\documentclass[%
 reprint,
superscriptaddress,
 amsmath,amssymb,
 aps,
]{revtex4-1}

\usepackage{graphicx}
\usepackage{dcolumn}
\usepackage{bm}
\usepackage{setspace}
\usepackage{xcolor}
\usepackage{soul}

\begin{document}

\preprint{APS/123-QED}

\title{A microrod optical-frequency reference in the ambient environment}

\author{Wei Zhang}
	\email{wei.zhang@nist.gov}
	\affiliation{National Institute of Standards and Technology, 325 Broadway, Boulder, Colorado 80305, USA \\}
\author{Fred Baynes}
	\affiliation{National Institute of Standards and Technology, 325 Broadway, Boulder, Colorado 80305, USA \\}
	\affiliation{Present address: School of Physical Sciences, The University of Adelaide, Adelaide, AU \\}
\author{Scott A. Diddams}
    \affiliation{National Institute of Standards and Technology, 325 Broadway, Boulder, Colorado 80305, USA \\}
\author{Scott B. Papp}
    \email{papp@colorado.edu}
    \affiliation{National Institute of Standards and Technology, 325 Broadway, Boulder, Colorado 80305, USA \\}

\date{\today}

\begin{abstract}
We present an ultrahigh-$Q$, solid-silica microrod resonator operated under ambient conditions that supports laser-fractional-frequency stabilization to the thermal-noise limit of $3 \times 10^{-13}$ and a linewidth of 62 Hz. We characterize the technical-noise mechanisms for laser stabilization, which contribute significantly less than thermal noise. With fiber photonics, we generate optical and microwave reference signals provided by the microrod modes and the free-spectral range, respectively. Our results suggest the future physical considerations for a miniature, low noise, and robust optical-frequency source.  

\end{abstract}

\pacs{Valid PACS appear here}
\maketitle


\textit{Introduction.}---Frequency-stabilized lasers based on evacuated, athermalized, vibration-isolated, and technical-noise-mitigated Fabry-Perot cavities define the state-of-art of frequency stability across the optical and microwave domains~\citep{Matei2017,Fortier2011}. These ultrastable lasers are critical scientific instruments for precision measurement science, such as atomic optical clocks ~\cite{Bloom2014,Hinkley2013}, gravitational wave detection~\cite{Rana2014}, very long baseline interferometry~\cite{VLBI} and other fundamental and applied research directions. An area of growing interest is to leverage the precision of optical cavities in challenging environments for geodesy~\cite{geo}, transportable optical lattice clock~\cite{PTBPRL2017}, and space-based fundamental physics tests ~\cite{SpaceCacciapuoti2009,Kolkowitz2016}, which will be enabled by miniature, robust and portable optical references ~\cite{LudlowOL07,MilloPRA2009,WebsterPRA2007,LeibrandtOE2011}.

Along the way of searching for miniature optical-frequency references based on whispering-gallery modes, work has focused on fluoride crystals ~\cite{SavchenkovJOSAB07,AlnisPRA2011,FescenkoMPQOE12,BaumgartelJPLOE12,Lim2017} that offer exceptionally high optical quality factors of $Q\sim10^{12}$. Previous research reveals that the fractional frequency stability (FFS) of a laser stabilized to crystals can reach $6 \times 10^{-14}$~\cite{AlnisPRA2011}, which requires complex ambient isolation, such as a vacuum chamber, multi-layer temperature control, and vibration isolation. Understanding the thermal-noise contribution in these devices has long been an important goal of analytical \cite{Lim2017,Matsko2007,GorodetskyJOSAB2004} and experimental work \cite{AlnisPRA2011,Lim2017}.

Fused-silica microrod resonators~\cite{ScottPRX2013,delhaye_laser-machined_2013,LohOptica2015} offer attractive properties, such as a solid monolithic structure, small optical mode volume, and use of silica material and fabrication properties. Though an alternative solution is a chip-based device, such as a high-$Q$ resonator based a spiral form~\cite{LeeNC2013} or external cavity semiconductor laser~\cite{LiangNC2015}.

\begin{figure}[t]
\centering
\includegraphics[width=0.85\linewidth]{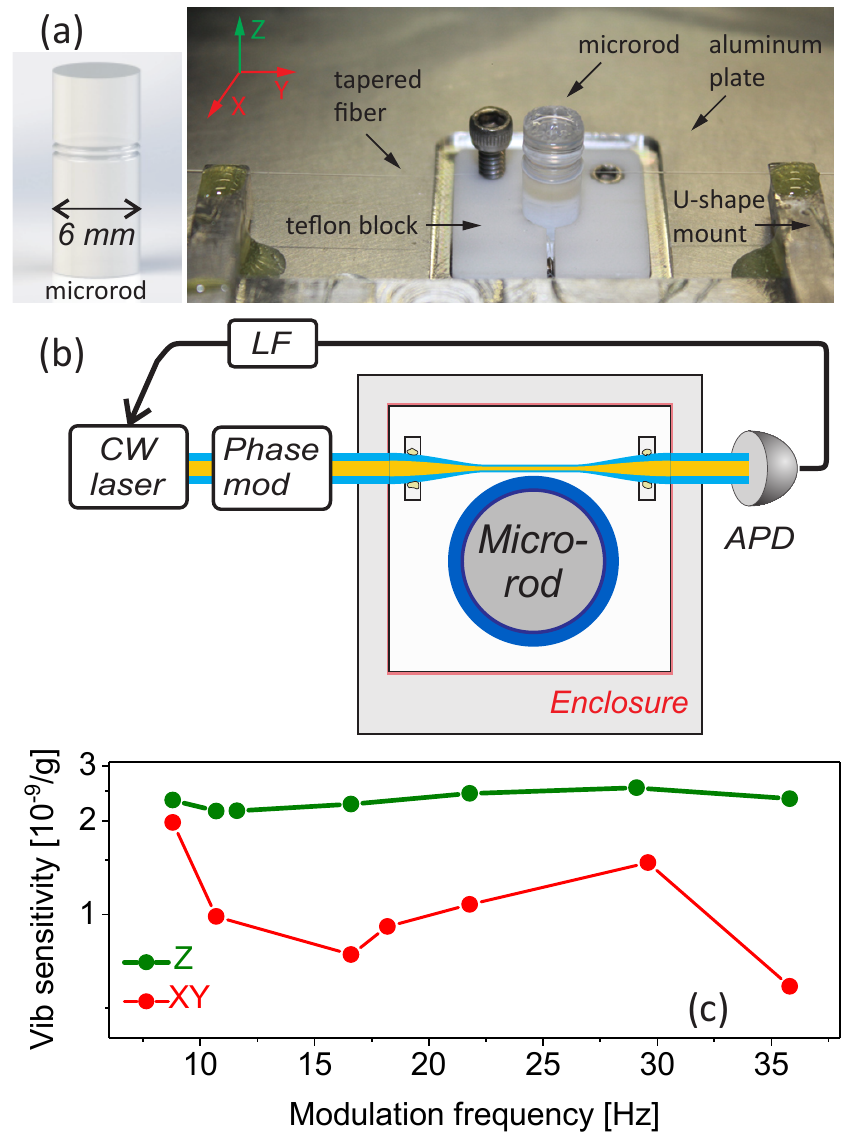}
\caption{(a) Photo of the microrod and the tapered fiber coupling. Inset: Drawing of the microrod. (b) Schematic of the microrod system and photonics components for laser stabilization. CW laser: continuous-wave laser; Phase mod: phase modulation; APD: avalanche photodetector; LF: loop filter. (c) Measurements of the fractional microrod resonance frequency vibration sensitivity.}
\vspace{-12pt}
\label{fig1}
\end{figure}

Here we report an optical- and microwave-frequency reference by frequency stabilization of a 1551 nm laser to a microrod resonator. The microrod is held in a heated aluminum enclosure with temperature control, but without vacuum or vibration isolation. After characterizing all technical noise sources in this stable laser system, we demonstrate that the fractional frequency stability reaches the thermal-noise floor at $3 \times 10^{-13}$ and the laser linewidth is 62 Hz. The thermorefractive (index of refraction) effect is the largest contribution. Furthermore, we generate an 11.8 GHz microwave signal by optical-to-microwave conversion based on stabilization to the microrod's free-spectral range; this microwave-generation procedure preserves the the fractional optical-frequency stability of the microrod for measurement intervals between 100 and 10,000 s.

The microrod is made of fused silica with a diameter of 6 mm. The resonator (Fig. 1a inset) with an unloaded $Q$ factor of 750 million is formed by CO$_2$ laser machining \cite{ScottPRX2013,delhaye_laser-machined_2013}. As shown in Fig. 1a, the microrod is held by a block of teflon, which is screwed on a aluminum plate (10 cm $\times$ 10 cm). The laser is coupled into the microrod by a tapered fiber glued on a U-shape mount, which is temporarily bolted on a translation stage. In construction of the setup, we performed a one-time adjustment of the tapered-fiber position for near critical coupling. After this optimization, the U-shape is released from the translation stage and glued on the aluminum plate. We attach a thermometer to the aluminum lid and wrap a hate tape on the aluminum enclosure to form a simple temperature-stabilized environment for the microrod.

As shown in Fig. 1b, an external cavity diode continuous-wave laser at 1551 nm is frequency-locked to the microrod with Pound-Drever-Hall (PDH) locking scheme~\cite{Drever1983}. We use a fiber-based waveguide electro-optic modulator (EOM) that provides a phase modulation at 8.1 MHz. By using a polarization-maintaining fiber coupler, 10\% of the EOM output power is coupled to a photodetector (PD) for residual amplitude modulation (RAM) detection~\cite{Wong85,Zhang2014}. The 90\% fiber coupler port, after an in-line isolator, is coupled into the microrod by a tapered fiber with a 60\% coupling efficiency. The transmission of the microrod is received by an avalanche photodetector (APD) to generate the PDH error signal by which the laser current modulation port is driven for a frequency lock; the feedback bandwidth is 500 kHz. In the entire setup, all components are either fiber-based devices or compatible with fiber in and output ports, allowing for a compact and robust system. We primarily characterize the microrod-stabilized laser by forming an optical heterodyne beatnote with a laser stabilized to a typical ultralow-expansion cavity ~\cite{Baynes15}. The frequency drift and noise of this beat signal are almost exclusively attributed to the microrod system. 

A focus of this paper is characterization of how the microrod-stabilized laser reacts to ambient conditions. One primary concern is vibration noise transferred to the microrod, which leads to deformation~\cite{ChenPhysRevA2006} and fluctuation of the resonance frequency. Since we do not use any passive or active vibration isolation, we rely on a low vibration sensitivity. Moreover, the vibration sensitivity of microresonator optical references has not been considered extensively, contrary to the case of Fabry-Perot cavities. To measure the vibration sensitivity, the microrod package is placed on an active isolation table, which is driven by the modulation signal from a vector signal analyzer. An accelerometer calibrates the motion of the isolation table and heterodyne beatnote is recorded for frequency response. The fractional frequency vibration sensitivity of the microrod is measured to be $2 \times 10^{-9}$/g (g=9.8~m/s$^2$) along the gravitational direction and $1 \times 10^{-9}$/g on the horizontal plane; see Fig. 1c. To analyze our data, we perform finite element analysis, which shows the cavity vibration sensitivity is between $3 \times 10^{-10}$/g along gravitational and $9 \times 10^{-10}$/g on horizontal plane. The difference between measurement and simulation is due to uncertainty in the geometry of the cavity and the cavity mounting. Note this microrod is an approximately cylinder shape, and our achieved vibration sensitivity relies on the small volume rather than vibration-immune design. When the whole system is resting on a fixed optical table, the ambient vibration is measured by accelerometer placed on top of the microrod package.

\begin{figure}[h!]
\centering
\includegraphics[width=\linewidth]{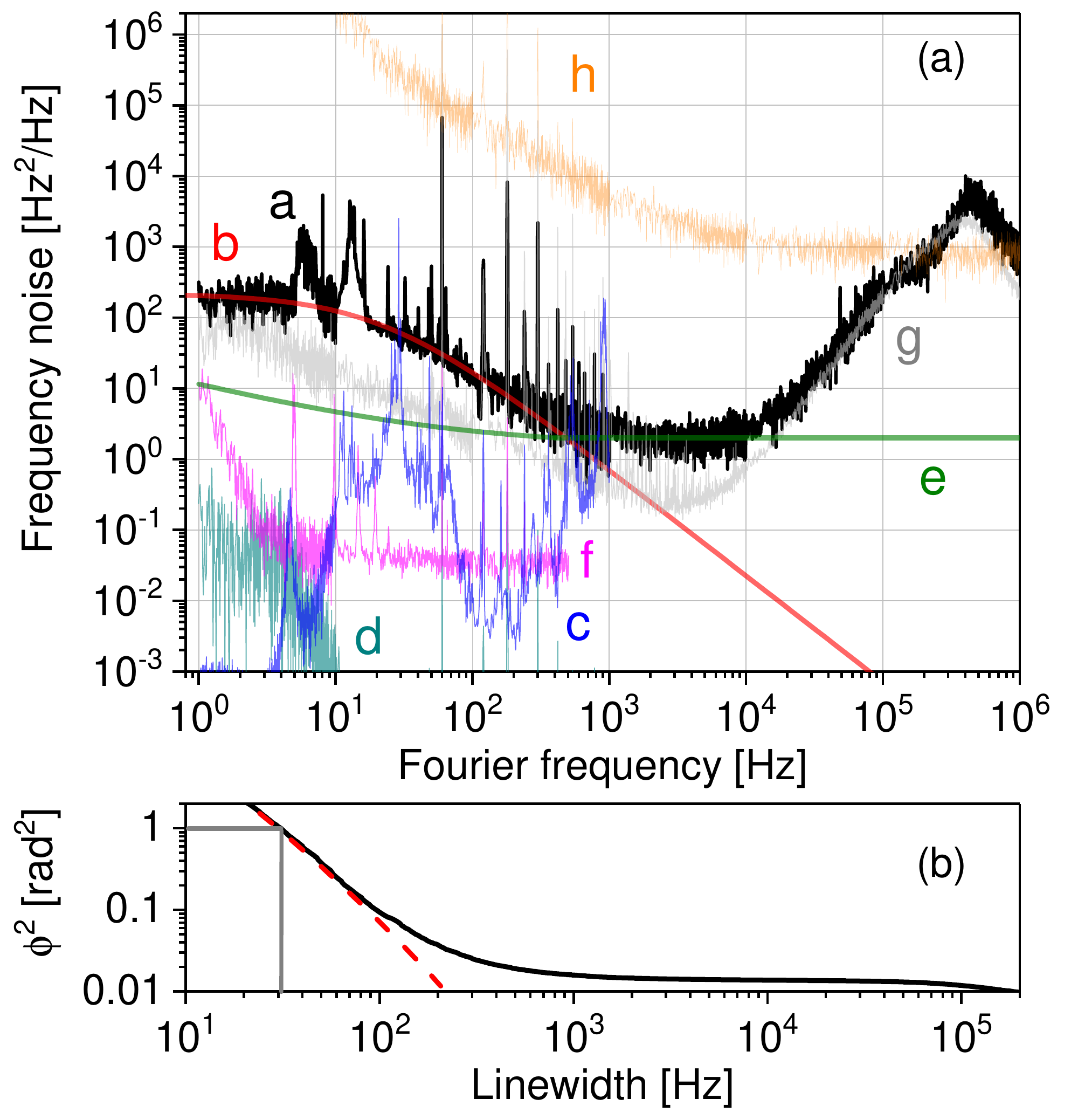}
\caption{(a) Frequency-noise power spectral density of microrod-stabilized laser and its noise contributions:
`a' the frequency noise of the heterodyne between the microrod laser and the reference laser;
`b' predicted thermal noise that dominates from 1 Hz to 300 Hz;
`c' vibration-induced frequency noise;
`d' intensity-induced frequency noise;
`e' PDH detector noise including shot noise and detector impedance noise;
`f' RAM-induced frequency noise;
`g' servo inloop error.
`h' the frequency noise of the free-running laser 
(b) The integrated phase noise up to 1 $\textrm{rad}^2$ leads to the laser linewidth of 62 Hz (black) and the limit due to thermal noise floor (red).}
\label{fig2}  \vspace{-9pt}
\end{figure}

Now we consider the optical-frequency noise of the microrod-stabilized laser, which is presented by the black `line a' in Fig. 2a along with a prediction of the thermorefractive-noise contribution (red `line b'). Our microrod laser is thermal-noise limited over decades in Fourier frequency, enabling a relatively narrow integrated linewidth. To understand this behavior, here we use a thermal model for silica material parameterized by the thermal-response time of the microrod. By solving the heat equation for a box and calculating the power-spectral density via the Wiener-Khinchin theorem, we find the expression $S_{\nu_M} \propto \nu_M^2 \, \alpha_n^2 \,/(\omega^2 \,+ \,\tau_i^{-2})$, where $\tau_i$ is the thermal time constant associated with the series of microrod thermal modes, $\nu_M$ is the optical frequency of a microrod mode, $\omega$ is angular frequency, and $\alpha_n$ is the thermorefractive coefficient. Following the important insight of Ref. \cite{Lim2017}, summation of all the thermal modes yields the power-law relationship $S_{\nu_M} \propto \nu_M^2 \, \alpha_n^2 \,/(1 + (\omega \, \tau_T)^{1.5})$. We normalize this expression to $\langle\delta T^2\rangle= k_B T^2/\rho C V$, where $\delta T$ is the temperature fluctuation associated with the heat capacity $C$, density $\rho$, and volume $V$ of the microrod's optical mode, and $T$ is the ambient temperature \cite{Matsko2007}. We measure the microrod's thermal time constant $\tau_T\approx0.01$ s by frequency-dependent heating the device with our laser. Our measurements establish the thermal-noise floor of microrod references and the path to improvements according to the straightforward formulas above.

We characterize other types of technical noise to understand how the microrod-stabilized laser behaves in ambient conditions. By applying the measured vibration sensitivity and the vibration power spectrum measured on top of the microrod package, the vibration-induced frequency noise is estimated and shown in Fig. 2a `line c,' which is below the predicted thermal noise (`line b') up to 400 Hz. The power fluctuation of the circulating light trapped in the microrod induces cavity resonance frequency fluctuation mainly due to light absorption. The transfer function from laser intensity to frequency is 60 kHz/$\mu$W at 1 Hz and 6 kHz/$\mu$W above 100 Hz.  We can apply a servo to stabilize the laser power and reduce the intensity-induced frequency noise; alternatively we lower the laser power to 1 $\mu$W at which microrod local heating is substantially reduced.  As shown in Fig. 2a `line d,' the intensity-induced frequency noise is below the predicted thermal noise floor. The use of low laser power elevates the contribution of the PDH detector noise (`line e').  Since the microrod has a relatively large cavity linewidth, RAM-induced frequency noise should be more substantial in which one part-per-million RAM corresponds to 0.8 Hz frequency fluctuation. The RAM-induced frequency noise (Fig. 2a `line f') is suppressed below thermal noise floor by inserting in-line isolation in the system and stabilizing the temperature of the EOM. 

A primary concern is ambient temperature fluctuations or drift that induce microrod frequency fluctuations. The microrod's thermal isolation is made of the aluminum plate and the teflon block shown in Fig. 1a. By applying a step change of the temperature on the enclosure and monitoring the laser frequency change, the time constant is measured to be approximately 1 minute. To estimate the temperature-induced frequency fluctuation on the microrod, the temperature fluctuation on the enclosure is multiplied with the transfer function from the enclosure to the microrod. Ambient temperature does not contribute significantly to the frequency-noise power spectrum, but we consider this effect in more detail below. 

Besides the ambient environment, technical limitations of the microrod's $Q$ factor and PDH detection are important to consider. Below 300-Hz Fourier frequency, the frequency noise is dominated by the predicted thermal-noise floor of the microrod. However, from 300 Hz to 10 kHz, shot noise and impedance noise of the APD shown as `line e' are the limit. `Line g' shows the bump from PDH servo above 10 kHz. At lower Fourier frequency, improving the inloop error requires a higher $Q$ resonator. Comparing to the frequency noise when laser is free running (`line h'), the microrod stabilization has an improvement by $10^{4}$ to the inloop level. The vibrations, laser intensity and RAM shown in Fig. 2a have been optimized and are not the main limitation. 

One manner to summarize the noise of the microrod-stabilized laser is the integrated phase noise from `line a' in Fig. 2a. The Fourier frequency equivalent to 1 $\textrm{rad}^2$ corresponds to a full-width at half-maximum (FWHM) linewidth of 62 Hz~\cite{arimondo_laser_1992}, as shown in Fig. 2b. This linewidth is consistent with the calculation according to the thermal-noise floor of `line b' in Fig. 2a.

\begin{figure}[h!]
\centering
\includegraphics[width=\linewidth]{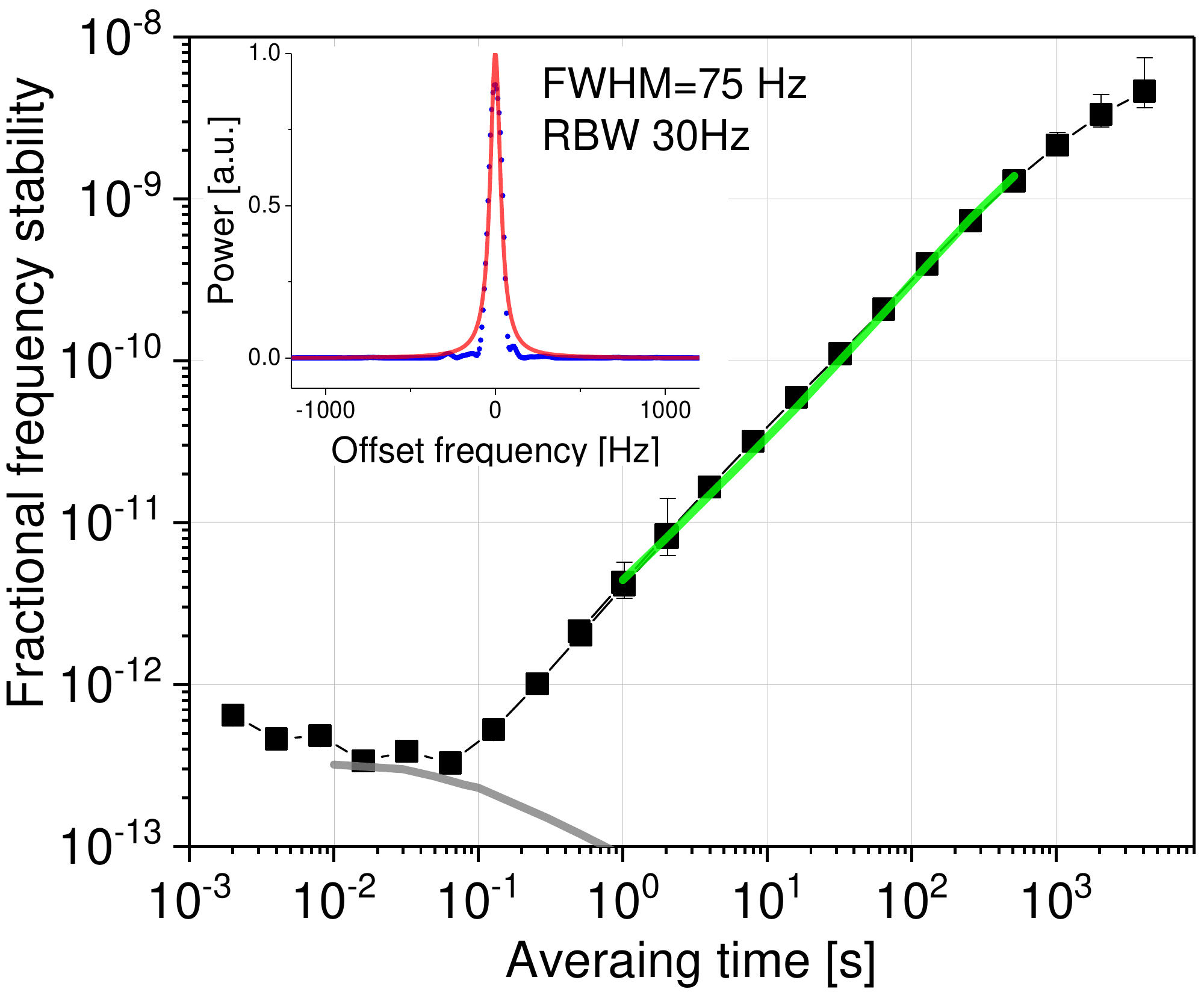}
\caption{FFS of the microrod-stabilized laser (black-square line), the thermal-noise floor (gray), and the temperature-induced noise (green). Inset: The measured linewidth of the heterodyne signal (blue dots) representing microrod-stabilized laser and the Lorentzian fit (red). }
\label{fig3} \vspace{-9pt}
\end{figure}


We also characterize the microrod-stabilized laser by an Allan-deviation analysis. We measure the heterodyne signal frequency with a triangle-type, dead-time-free frequency counter and calculate the Allan deviation. We use counter gate times of 2 ms and 500 ms to increase dynamic range. As shown in Fig. 3, for averaging time $0.01\,\text{s}<\tau<0.1\,\text{s}$, the FFS reaches the thermal noise floor (gray line) at $3 \times 10^{-13}$. For $\tau>1 \,\text{s}$, the FFS increases due to frequency drift caused by temperature fluctuations on the enclosure. 
Moreover, the full-width half-maximum linewidth of the heterodyne signal (inset of Fig. 3) as determined with an RF spectrum analyzer is 75 Hz (Lorentz fit, 30 Hz resolution bandwidth), which is fully consistent with frequency noise and Allen deviation results.

\begin{figure}[t]
\centering
\includegraphics[width=\linewidth]{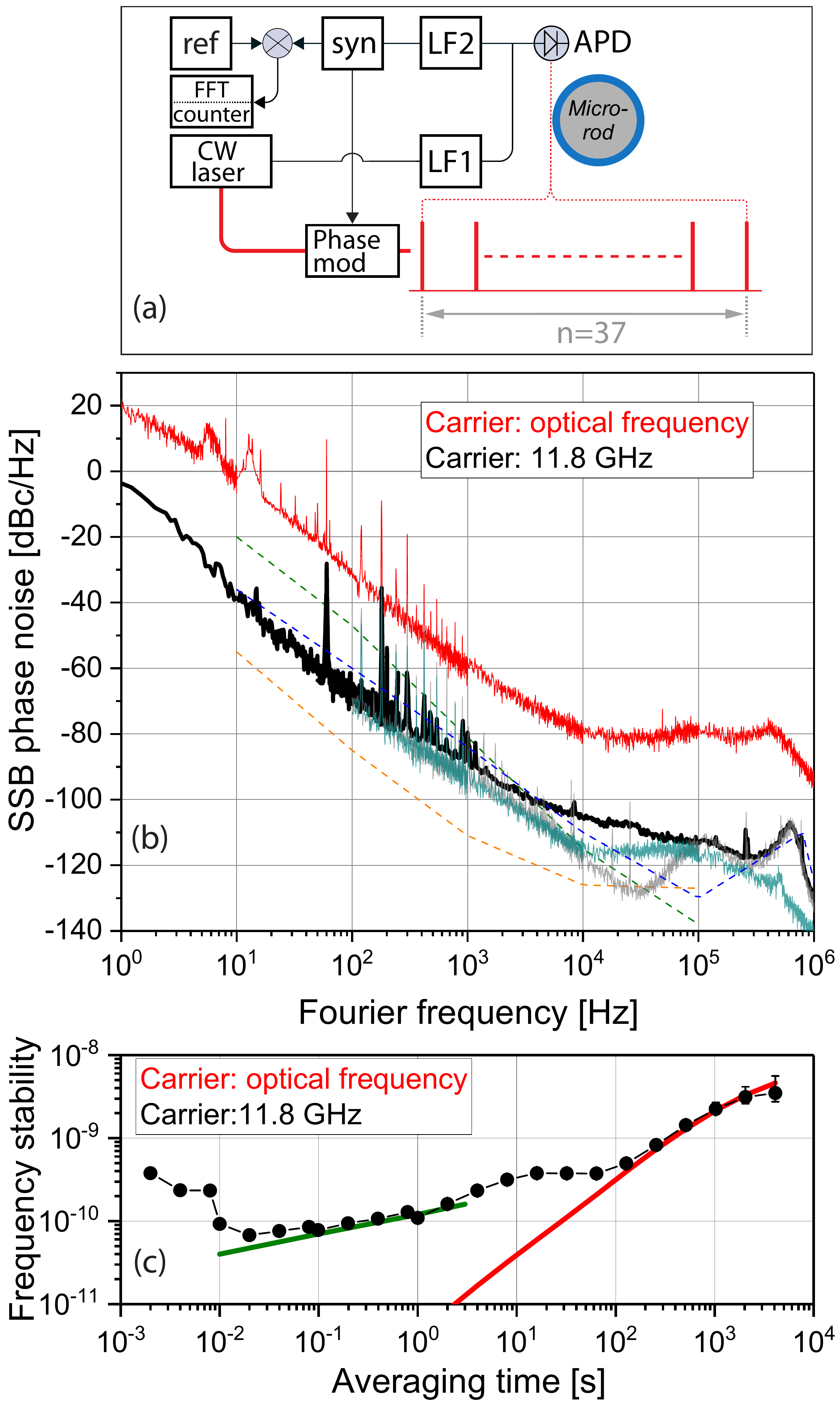}
\caption{(a) Schematic of the microrod-stabilized microwave oscillator. CW laser, continuous-wave laser; Phase mod, phase modulator; APD, avalanche photodetector; syn, microwave synthesizer; LF, loop filter;  ref, reference microwave synthesizer; FFT, fast Fourier transform analyzer. (b) Phase noise of the 11.8-GHz microrod-oscillator (black) is lower than the microrod-stabilized laser (red), and is predominantly limited by inloop errors (green and gray) of the PDH locks. Below 30 Hz the data is converted from a time-domain measurement. For comparison, Synergy DRO-100 (green-dashed), Wenzel MXO-FR (orange-dashed) and frequency division of Brillouin lasers~\cite{LiScience2014} (blue-dashed) are presented. (c) FFS of the microrod oscillator (black), compared with the inloop errors (green) and the FFS of the microrod laser (red).}
\label{fig:4} \vspace{-18pt}
\end{figure}

With the optical-frequency noise properties of the microrod-stabilized laser established, we turn to generating a low-phase-noise microrod-stabilized oscillator. Here the concept is to stabilize a microwave oscillator to the $11.8$-GHz FSR, using our laser and PDH locking. The optical frequency of the microrod mode is intrinsically linked by momentum conservation to the free-spectral range (FSR) through $\nu_M = \textrm{FSR}\times M$, although the FSR is sensitive to myriad physical parameters, including wavelength, temperature, pressure, resonator shape, index of refraction, and electromagnetic fields. We are interested to explore the phase noise, lower-bounded by $S_\textrm{FSR} = S_{\nu_M}/M^2$, and the Allan deviation, lower-bounded by $\delta(\textrm{FSR})/\textrm{FSR}=\delta(\nu_M) / \nu_M$ of a microwave oscillator that is locked to the FSR, where $\delta()$ indicates the fluctuations of the quantity ~\cite{Maleki2011}. In these limits the microwave oscillator would offer the same FFS as our microrod-stabilized laser. For stabilization to the microrod FSR, we use an optical comb $\nu_n = \nu+ n \, f_m$, generated from our laser frequency $\nu$ and a phase modulator driven by an oscillator at $f_m$. The laser and comb mode $n$ are PDH-locked to respective microrod modes belonging to the same family; the inloop errors of these locks are $S_{\nu_0}$ and $S_{\nu_n}$, respectively~\cite{SwannOE2011}. Therefore we can approximate the microwave oscillator noise as $S_{f_m}=S_\textrm{FSR} + \left(S_{\nu_0} + S_{\nu_n}\right)/n^2$.

In experiments, 10\% laser power output is locked to the microrod by PDH, and 90\% is sent to a phase modulator, driven by a microwave synthesizer, to generate comb lines; see Fig. 4a. We choose $f_m$ at 11.8 GHz, and the $n=37$ comb line is locked to the microrod simultaneously by the second PDH in which the frequency-modulation port of the microwave synthesizer is used as the actuator. Frequency multiplication also facilitates locking the synthesizer. After the two comb modes are locked to the one cavity, the microrod-stabilized oscillator is measured by comparing with a reference microwave synthesizer (Agilent E8257), which is locked to a hydrogen maser. A phase-noise analyzer and frequency counter are used for the comparison.

The black trace in Fig. 4b shows the phase noise of the 11.8 GHz microrod-stabilized oscillator, which is $-40$ dBc/Hz at 10 Hz, $-70$ dBc/Hz at 100 Hz, and $-120$ dBc/Hz up to 1 MHz. The data is largely explained by the inloop errors $S_{\nu_0}/n^2$ (green trace) and $S_{\nu_n}/n^2$ (gray trace). This level of performance is somewhat comparable to other compact oscillators,  at the present level of development of our microrod system. Comparing the microrod-stabilized laser at optical frequencies (red trace) to the microrod-stabilized oscillator shows phase noise that is reduced by slightly more than $20\,\log{(37)}$ at some Fourier frequencies, as we would expect from $S_{f_m}$ when the inloop error terms are insignificant.

The solid points (red trace) in Fig. 4c show the measured FFS of the microrod-stabilized oscillator (laser). These data are acquired with zero-dead-time frequency-counter measurements and Allan deviation analysis. While the oscillator FFS is limited by the previously described inloop error (green trace) in the millisecond range, for measurement intervals between 100 and 10,000 seconds the FFS is comparable in the microwave and optical domains. We expect this behavior given the link $\nu_M = \textrm{FSR}\times M$, and we can observe it due to microrod drift. Apparently, other fluctuations of the ambient environment are insignificant in this timescale.

In conclusion, we demonstrate an optical-frequency reference in the ambient environment based on a microrod-stabilized laser with a FFS at the thermal-noise-limit of $3 \times 10^{-13}$.  We characterize the technical noise, which are below the thermal noise. The long-term stability is dominated by temperature instability, which can be optimized by referencing laser to atomic transition in microfabricated rubidium  cell~\cite{loh_microresonator_2016_OE}. Furthermore, we show that microrod optical stabilization may also be applied for microwave-signal generation. Our 11.8-GHz microrod-stabilized oscillator provides competitive overall performance, and it explores the relative physical influences of optical resonances and the FSR.

We thank D. Nicolodi and L. Stern for their comments. This work was supported by NIST and DARPA. NIST does not seek copyright of this work. Product names are given for information only.

\bibliography{reference.bib} 

\end{document}